\documentstyle [epsf]{mn}
\def\be{\begin{equation}}
\def\ee{\end{equation}}
\def\ba{\begin{eqnarray}}
\def\ea{\end{eqnarray}}
\def\rms{{\em rms }}
\def\ps{{P\&S }}

\begin{document}

\title[Cluster Satellites]
{\bf The Rise and Fall of Satellites in Galaxy Clusters}
\author[G.Tormen]
{Giuseppe Tormen \\
Max Planck Institute f\"{u}r Astrophysik,
Karl-Schwarzschild-Strasse 1, 85740 Garching bei M\"{u}nchen - GERMANY
and\\
Institute of Astronomy, University of Cambridge, Madingley Road, 
Cambridge CB3 0HA - UK\\
\smallskip
Email: bepi@mpa-garching.mpg.de}

\date{MNRAS {\bf290,} 411-421 (1997)}
\pubyear{1997}

\maketitle

\begin{abstract}
We use $N$-body simulations to study the infall of dark matter 
haloes onto rich clusters of galaxies. After identification of 
all cluster progenitors in the simulations, we select those haloes
which accrete directly onto the main cluster progenitor. 
We construct the mass function of these merging {\em satellites}, 
and calculate the main orbital parameters for the accreted lumps. 
The average circularity of the orbits is $\epsilon \simeq 0.5$,
while either radial or almost circular orbits are equally avoided.
More massive satellites move along slightly more eccentric orbits,
with lower specific angular momentum and a smaller pericentre.
We find that the infall of satellites onto the main cluster 
progenitor has a very anisotropic distribution. This anisotropy
is to a large extent responsible for the shape and orientation
of the final cluster and of its velocity ellipsoid.
At the end of the simulations, the major axis of the cluster 
is aligned both with that of its velocity ellipsoid, and with the 
major axis of the ellipsoid defined by the satellite infall pattern, 
to $\approx 30\degr$ on average.
We also find that, in lower mass clusters, a higher fraction 
of the final virial mass is provided by small, dense satellites. 
These sink to the centre of the parent cluster and so enhance 
its central density. This mechanism is found to be partially 
responsible for the correlation between halo masses and 
characteristic overdensities, recently highlighted by 
Navarro, Frenk \& White (1996). 
\end{abstract}


\section{Introduction}\label{sec:intro}

In cosmological models of gravitational instability, structures
form by the collapse of small perturbations of some initial 
density field. The most successful flavour of these models is 
the hierarchical clustering picture, where matter clusters on small 
scales first, and structures form in a bottom-up fashion. 
In this scenario, clusters of galaxies represent the most 
recently assembled structures, and are therefore dynamically
young systems. For this reason, they are ideal candidates to study
the link between the final structure and morphology of an object
and its formation process. In this context, it is of great interest 
to investigate in some detail the accretion of matter within the 
proto-cluster. The study of the recursive merging of smaller size 
objects onto the main cluster progenitor is a powerful tool to unveil 
the cluster dynamical history, and can possibly help us to understand 
some details of its final structure.

A detailed investigation of the cluster merging history usually
requires either Monte--Carlo techniques or numerical simulations.
In the former case, one uses an analytical model for the clustering 
of structures, like the Press \& Schechter (1974) (hereafter \ps)
formalism, and extensions (Bond et al. 1991; Bower 1991; Lacey \& Cole
1993; Lacey \& Cole 1994), to produce Monte Carlo realizations of the 
merging histories of dark matter haloes. 
These in turn are the starting point of a recipe for galaxy formation,
whose predictions can be compared to the observed properties of real 
galaxies and clusters. 
Extensive work in this direction has been done in the past few years 
(Kauffmann \& White 1993; Cole et al. 1994; Heyl et al. 1995; 
Kauffmann 1995), in an attempt to interpret the various observations 
in a unified and global framework.
While this work provides a plausible link between the properties
of dark matter haloes and those of galaxies and clusters, it is 
necessary to complement the investigation with numerical simulations. 
These, although limited to a smaller number of applications, do not 
need as many simplifying assumptions on the details of structure 
formation. Therefore one can address issues not accessible to Monte 
Carlo techniques, as the dynamical equilibrium of clusters, the study
of their density and velocity profiles, the characterization of the
orbits of infalling haloes, the survival times of haloes after merging
with the main cluster, and others.
These results are also useful to improve the Monte Carlo methods, 
as they can tell which assumptions are more realistic for the galaxy 
formation recipe.

Although many numerical simulations of cluster formation have been 
performed in the past (White 1976; Quinn, Salmon \& Zurek 1986;
Navarro, Frenk \& White 1995; Tormen, Bouchet \& White 1997)
there is little work on the connection between the details of the
cluster merging history and the final cluster configuration,
in a cosmological context. One reason for this is that such analysis
requires simulations with very high mass and force resolution,
as only recently have become available.
In the present paper we present the results of such a study. 
Using a sample of nine dark matter haloes of rich galaxy clusters,
obtained from high resolution $N$-body simulations, we identify the 
merging history of all cluster progenitors, and extract from that 
a population of {\em satellites}, i.e. progenitor haloes accreting 
directly onto the most massive cluster progenitor at any time.
We study this population and characterize its dynamical properties. 
We then use the information on the infalling satellites as a key to
interpret the cluster formation history, and its final shape and 
structure.
In a forthcoming paper (Tormen 1997, in preparation) we shall make 
a detailed comparison of the merging history of the clusters with 
the predictions of the extended \ps formalism.

The outline of the paper is as follows. Section 2 briefly presents
the simulations, and describes the method used to define their
halo population. In Section 3 we study the cluster satellites. 
We look at the way they merge with the main halo, and measure 
their mass distribution. Then, by moving the satellites in the 
spherically averaged, static potential of the main halo, as 
measured at each time output, we integrate their orbits and 
derive the main orbital parameters.
In Section 4 we study the connection between shapes and orientations 
of clusters and the infall pattern of their satellites, and 
investigate the dynamical origin of universal dark matter density 
profiles (Navarro, Frenk \& White 1996).
Finally, in Section 5 we summarize the paper and present the main 
conclusions.

\section{Method}\label{sec:method}

\subsection{Simulation setup}\label{sec:sim}

For the present analysis we use the $N$-body simulations presented
in Tormen et al. (1997), where complete details on the
simulations may be found. In summary, our sample consists of nine 
dark matter haloes of galaxy clusters, with an average mass of 
$1.13 \times 10^{15} M_{\sun}$, resolved by $\approx 20000$ dark
matter particles each within the virial radius, with an effective 
force resolution of $\sim 25$ kpc. 
The haloes come from an Einstein--de Sitter universe
with scale-free power spectrum of fluctuations $P(k) \propto k^n$, 
and a spectral index $n=-1$, the appropriate value to mimic a
standard cold dark matter spectrum on scales relevant to cluster
formation. The simulations have a Hubble parameter
$H_0 = $50 km s$^{-1}$ Mpc$^{-1}$ and are normalized 
to match the observed local abundance of galaxy clusters 
(White et al. 1993).
Each simulation was performed using two sets of particles, a first 
high resolution set to actually form the cluster, and a low 
resolution set of massive background particles to model the
large-scale tidal field of the simulation out to a scale $L = 150$ Mpc.
For each of the nine simulations we shall use ten outputs, ranging from 
$z \simeq 6$ to $z = 0$, for a total of 90 outputs.

\subsection{Identification of dark matter lumps}\label{sec:hid}

The relatively low number of particles of each simulation 
($N \approx 60000$ on average) allowed us to use the potential energy 
of particles to identify the location of dark matter lumps 
(Efstathiou et al. 1988).
We defined lumps by an overdensity criterion, and included all particles 
within a sphere of mean overdensity $\delta_v = 178$, centred on the 
particle with lowest potential energy. The value $\delta_v$ corresponds
to the virial overdensity in the model of a spherical top-hat collapse.
We shall call virial radius $r_v$ the corresponding radius of the lump.
We identified all lumps with at least five particles within $r_v$.

In order to determine the minimum size of lumps to use in our analysis,
we performed the following test. 
We selected, for each output of each simulation, all lumps formed
by $n$ particles, and looked at the fate of those particles in the 
next time output. The particles of a lump will now be part of 
other lumps. We took note of the largest fraction of lump relic
that was found in a single halo, and looked at the distribution
of such quantity. If all, or most, of a lump is found in another halo
on the next time output, the lump is a dynamically consistent object.
We performed the analysis for several values of $n$, from $n=5$ up, 
and found convergence in the resulting main relic distribution for 
lumps with at least 8 particles.
For such lumps, in 95 per cent of the cases the largest relic carried 
at least half of the original mass, and in 70 per cent of the cases it carried 
more than 80 per cent. On the other hand, almost 20 per cent of lumps with 5 
particles were completely dissolved in the field by the next output. 
Therefore we chose to limit our analysis to lumps with at least 
eight particles within their virial radius. All remaining particles,
including those in smaller lumps, will be generally referred to as 
{\em field particles}.
Since we limit our analysis to lumps which are progenitors of the
nine simulated clusters, these lumps are only made up with high 
resolution particles.

\section{Merging of Satellites}\label{sec:msat}

Let us call, at any given redshift, {\em progenitors} of a cluster 
all haloes containing at least one particle that by $z=0$ will be
part of the cluster, and also all field particles which will end up
in the final cluster. At any redshift, we define the {\em main}
(or {\em largest}) {\em progenitor} of a cluster as the progenitor 
halo containing the largest number of particles which will be
part of the final cluster. In all cases this coincided with the most
massive progenitor halo, although, with the present definition, the 
main progenitor in a simulation is not necessarily the most massive one.
To help clarifying the meaning of these terms, one can refer to the
schematic merging history of halos given by the merging tree in Fig.6 of 
Lacey \& Cole (1993).

We termed {\em satellites} those progenitor lumps which were directly 
accreted onto the main cluster progenitor. Operationally, we started 
from the list of progenitors at a given output of the simulation, and 
selected those haloes for which at least one particle was found in the
largest cluster progenitor in the next time output of the simulation. 
We shall characterize these satellites by studying their mass 
distribution, and by calculating their orbital parameters.
We shall call {\em identification time} $t_{id}$ of a satellite the 
last output time before the satellite accretes onto the main cluster 
progenitor. Since we consider ten outputs for each simulation, there
are nine outputs (all but the last) to identify satellites.

\begin{figure}
\epsfxsize=\hsize\epsffile{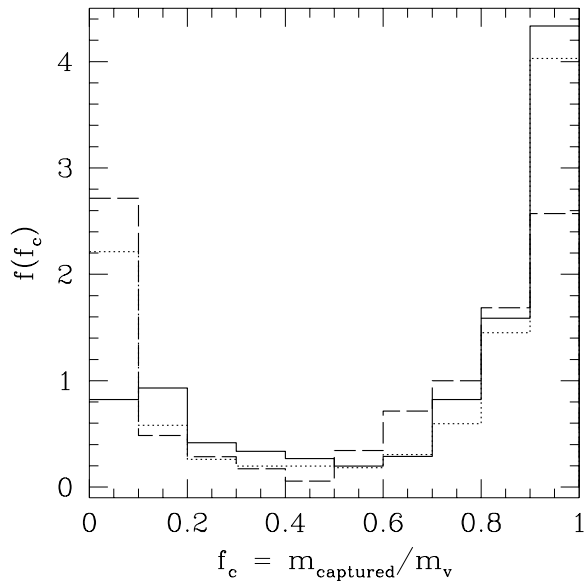}
\caption{Number weighted distribution of the fraction of satellite mass
which is captured by the main cluster progenitor.
Each curve refers to a different range of particle number for the 
satellites: $m_v <20$ particles (solid), $20 < m_v < 100$
particles (dotted) and $m_v > 100$ particles (dashed).
The two extreme situations are most common: satellites stripped 
by a small amount of their mass, with $m_{captured}/m_v < 0.2$, 
and satellites captured (almost) entirely, with 
$m_{captured}/m_v > 0.8$. Intermediate cases are very rare.}
\label{fig:fmer}
\end{figure}

In the hierarchical clustering picture of structure formation, 
matter clusters on small scales first, and dark matter haloes of 
a given size are formed by the assembly of smaller haloes which 
were formed earlier. In this idealized model, when a satellite 
lump is accreted onto a more massive halo, all of the satellite 
mass is incorporated in the bigger object. Although this description 
is fairly close to what we observe in $N$-body simulations of 
hierarchical clustering, the actual collapse of structures is more 
complicated than its idealized version; in particular, satellites 
may be tidally stripped by the accreting halo, so that only a 
fraction of the satellite mass becomes part of the accreting system. 
This is clearly illustrated in Fig.~\ref{fig:fmer}, where we show
the number distribution of the fraction $f_c$ of satellite mass
captured by the main cluster progenitor, for all time outputs and 
all clusters together.
Each curve refers to satellites with different number of particles.
We observe that most of the satellites transfer either very little 
or most of their mass to the main cluster progenitor, whereas less 
than 20 per cent of the satellites have $f_c \in [0.2, 0.8]$. 
This trend is more evident for more massive satellites.
While this result may be partially due to a few satellites which, 
by chance, happen to cross the virial radius of the main progenitor
at the time of a simulation output, it is more likely that
the distribution reflects different physical situations.
We shall address the issue at the end of the section on the 
orbital parameters.

\subsection{Mass function of satellites}\label{sec:mmer}

It is interesting to look at the spectrum of masses that clumped 
together and formed the clusters in our sample. This will tell 
us, for example, how much of a cluster's mass was made up by violent 
encounters with massive satellites, as opposed to smooth accretion 
of small sized systems. We want to select from the satellite list 
those who significantly contribute with their mass to the formation 
of the cluster. We do this by eliminating all satellites captured by
less than 50 per cent in mass, that is with $f_c < 0.5$. 
This choice seems reasonable, as Fig.~\ref{fig:fmer} shows that 
there is a clear separation between 
satellites barely stripped and satellites really captured.
Moreover, with this distinction, a satellite excluded from
the selection in one output may be included in the next, if 
it is really captured by the main halo.
We shall call {\em merging time} of the satellite onto the cluster 
the time when the satellite first crosses the virial radius of 
the main cluster progenitor. Since the separation between consecutive
outputs of our simulations is $\Delta t \simeq 1.6$ Gyr, 
we can only bracket the estimated merging time of a satellite 
between the identification time $t_{id}$ and the next output 
of the simulation at $t_{id} + \Delta t$.

\begin{figure}
\epsfxsize=\hsize\epsffile{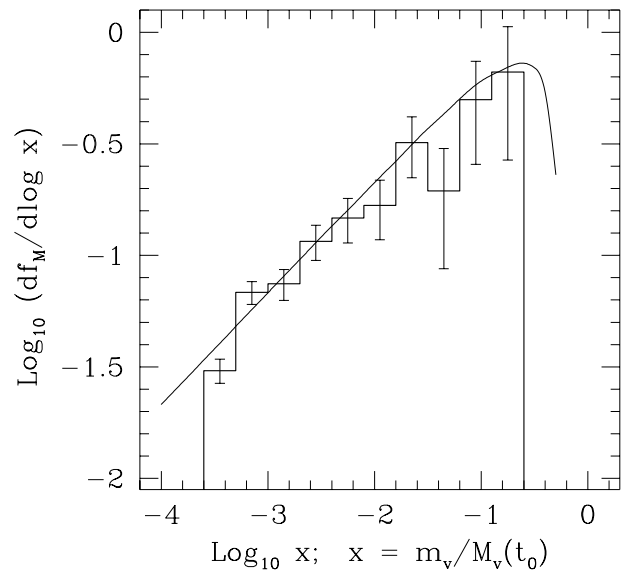}
\caption{Mass function of satellites. The histogram shows the
fraction of cluster mass accreted in satellites of mass $m_v/M_v(t_0)$, 
averaged over all clusters in our sample. Only haloes accreted 
after the cluster formation redshift $z_f$ (the redshift when 
the most massive progenitor reaches half of the final cluster 
mass), and captured by at least 50 per cent of their virial mass, 
are considered.
The mass of satellites is given in units of the {\em final} 
cluster mass; error bars are Poissonian. The solid curve is a 
prediction from Monte Carlo merging trees, taken from 
Lacey \& Cole 1993 (their Fig.~12).}
\label{fig:mmer0}
\end{figure}

Fig.~\ref{fig:mmer0} shows the mass distribution of satellites
captured by at least 50 per cent in mass; the abscissa is the virial
(i.e. total) mass of the satellite normalized to the final
mass of the cluster.
The smooth curve is taken from Lacey \& Cole 1993, and is the 
expected mass distribution of satellites obtained from Monte--Carlo 
merging histories of haloes. The prediction does not depend on the 
final cluster mass if one considers only the accretion events 
happening after the {\em formation redshift} $z_f$, defined as
the redshift when the most massive cluster progenitor first reaches
at least half of the final cluster mass (Lacey \& Cole 1993, 1994). 
In our case, we only considered satellites with $z_{id} \leq z_f$.
To give an idea of our statistics, there are roughly 1400 satellites
merging with the clusters, of which almost 600 identified after
$z_f$.
The histogram is normalized to the fraction of cluster mass provided 
by the satellites with $f_c \ge 0.5$. This is done to keep account 
of the matter accreted onto the main cluster progenitor in both 
field particles and satellites with $f_c<0.5$.
The agreement between model and simulation data is very good, as the 
curve matches the histogram over nearly three orders of magnitude in 
mass.

\begin{figure}
\epsfxsize=\hsize\epsffile{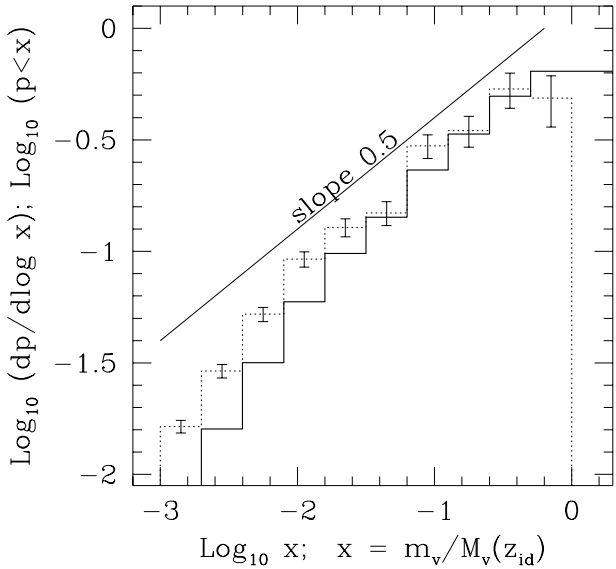}
\caption{The same distribution shown in Fig.~\ref{fig:mmer0} is
shown here with satellite masses normalized to the mass of the
main cluster progenitor at the time of capture. 
Both the cumulative (solid line) and the differential 
(thin dotted line, with Poissonian error bars) are drawn.
These curves show the relative occurrence of merging events 
with satellites of a given mass ratio $m_v(z)/M_v(z)$.}
\label{fig:mmer1}
\end{figure}

In Fig.~\ref{fig:mmer0}, the $x-$axis has masses normalized to
the final mass of the clusters, for the purpose of comparing it 
with the Monte--Carlo prediction. However, it is more 
instructive to look at the same distribution normalizing masses 
to that of the main cluster progenitor at the identification time 
of the satellite, $t_{id}$. This is done in Fig.~\ref{fig:mmer1}, 
where the differential (dotted histogram) and cumulative (solid 
histogram) mass functions of the accreted satellites are plotted. 
Now we use all satellites identified at any $t_{id}$, but always 
with $f_c \ge 0.5$.
The (Poissonian) error bars refer to the differential distribution. 

In the mass range $\log x \ga -2$ the distribution follows a power 
law with slope $0.5$, the same slope of the distribution in 
Fig.~\ref{fig:mmer0}.
Normalization is the same described above, showing that, on average,
one third of the cluster mass was accreted either in field particles
or in satellites captured by less than half of their mass.
The next largest contribution, about 20 per cent of the final cluster mass, 
comes from encounters with satellites with mass ratio $m_v/M_v \geq 0.5$. 
On the whole, roughly $40--45$ per cent of the total cluster mass comes from 
encounters with satellites with $m_v/M_v \geq 0.1$. These figures
are quite robust, as they remain essentially the same if we use in
the histograms all satellites (regardless to the captured fraction 
of their mass) or if we use for the satellite mass the actual captured 
mass instead of the virial mass.
How frequent are encounters with massive satellites? 
Let us consider only satellites with $f_c \geq 0.5$.
The average number per cluster of encounters with satellites 
with mass ratio $\geq 0.1$ ($\geq 0.2; \geq 0.5$ respectively) is
$2.7 \pm 0.5$ ($1.5 \pm 0.4$, $0.1 \pm 0.1$) from the time $t_f$ when
a cluster reached half of its final mass ($z_f \approx 0.5$ on average)
to the present time, and $13.7 \pm 1.4$ ($7.4 \pm 1.0$, $2.3 \pm 0.5$)
over a Hubble time. Error bars are $1\sigma$ estimates from
bootstrap resampling over the clusters.
It has been shown (Tormen et al. 1997) that encounters with satellites
with mass ratio larger than roughly $0.2$ can perturb the \rms
velocity of the cluster by as much as 30 per cent within the virial radius. 
According to the present result, a rich cluster should undergo such 
encounters between one and two times since $z_f$.

\subsection{Orbital parameters}\label{sec:orb}

In this section we study the orbits of the satellites which accrete 
directly on the main cluster progenitor. Since we want to investigate
how satellites fall onto the cluster, we are interested in the
{\em initial} orbital parameters, which define the orbits {\em prior} to 
the actual merging. 
Of course, after a satellite crosses the virial radius of the main 
cluster, different dynamical processes (tidal stripping, dynamical 
friction, encounters with other halos, etc...) become more relevant, 
and its orbit will be modified. 
However, the {\em initial} orbits are still very interesting as they 
trace the accretion process. Moreover, they can be used as an input
for a more realistic modelling of the dynamical processes listed above,
for example in semianalitic models of galaxy formation.

In order to integrate the orbits of satellites we used the static 
spherical potential approximation (e.g. Binney \& Tremaine 1987,
their Eq.(3.13)), and moved the satellites in the potential 
of the main cluster progenitor as measured at time $t_{id}$. 
The radial potential profile $\Phi(r)$ was obtained by averaging the 
actual potential of all the particles in each output in
spherical shells of equal logarithmic width 0.05, centred on the 
main cluster progenitor. A further Gaussian filter, of logarithmic
width $0.2$, was applied to the profiles to make them smoother and so
avoid multiple solutions in the orbital parameters.
Although we calculated the orbital parameters for all satellites,
here we shall only show results for satellites with $f_c\geq 0.5$.

The approximations made will lead sometimes to unrealistic solutions,
especially {\em (a)} for massive satellites, whose contribution
to the total potential is not negligible, and {\em (b)} at early
times, when the main cluster progenitor grows faster, and so the
static approximation is poorer, and when there is more chance of
having big neighbouring haloes that invalidate the spherical 
approximation.
In order to minimize these problems, in this Section we limit our 
analysis to outputs where the main cluster progenitor has at least 
half of the cluster final mass, that is to redshifts $z_{id} < z_f$.
With this choice, the mass of satellites, relative to that of the 
main cluster, is kept small (in our case $m_v/M_v <0.25$ for more 
than 99 per cent of the satellites), and the static model is closer to 
reality, since the potential well of the cluster has already been 
built up. This selection will somehow bias our statistics, because 
we automatically exclude massive ($m_v/M_v \ga 0.25$) satellites
from the analysis. However, we prefer this choice to the 
alternative of including unreliable satellites in the results.

In order to compare results from different outputs and from different 
clusters, we need to rescale them to uniform units. In particular,
lengths should be expressed in units of the virial radius of the 
main cluster progenitor.
Now, even in a static potential, the virial radius $R_v$ of the 
main halo will grow with time, since the mean background density
of the universe is decreasing like $(1 + z)^3$. As a consequence, 
from the time $t_{id}$ when we identify a satellite, and the later 
times $t_v$, $t_p$ when the satellite crosses the virial
radius and reaches the orbit pericentre, $R_v$ will have grown,
sometimes by a not negligible amount.
This growth must be taken into account since our satellites are 
identified at different distances from the main progenitor, and so
cross its virial radius at different times. 
For every satellite, we calculate $t_v$ and $R_v(t_v)$
moving the satellite in the static potential of the cluster, 
and moving the cluster virial radius outward until the satellite
crosses it.

A proof that the assumed model is a better approximation to the
real clusters at late than at early times comes from the 
distribution of orbit pericentres: the most obvious unphysical 
result is a satellite not merging at all with the main cluster, 
that is, with a pericentre $r_p > R_v$. This actually
happens for 14 per cent of the satellites with $z_{id} \geq z_f$,
but only for 2 per cent of the satellites with $z_{id} < z_f$. 

\begin{figure}
\epsfxsize=\hsize\epsffile{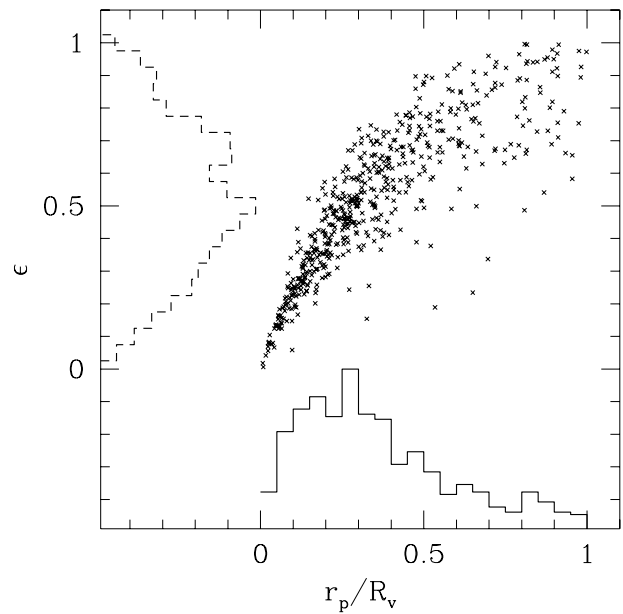}
\caption{Scatter plot of the orbital pericentre $r_p/R_v$
and the orbital circularity $\epsilon$. The two histograms
show the corresponding projected distributions (solid line for
$r_p/R_v$ and dashed line for $\epsilon$).}
\label{fig:orb1}
\end{figure}

We must also be careful that the satellites are not too far from 
the main cluster progenitor at their identification time. In fact, 
at large distances from the cluster, orbits are more easily 
influenced by neighbouring massive objects, and the spherical 
average of the potential is no more a good description of the 
true field. Fortunately, in practice this is not a problem, since 
all selected satellites have $r_{id} < 3 R_v(t_{id})$, and 90 per cent 
have $r_{id} < 2 R_v(t_{id})$.

What is the shape of the orbits? How close do the satellites come 
to the centre of the main cluster? How much angular momentum do they 
carry? These are some questions we want to answer in this Section.
Fig.~\ref{fig:orb1} shows a scatter plot of the orbital
circularity $\epsilon$ versus the pericentre of the satellite's
orbit (the latter rescaled to the virial radius of the main cluster
progenitor, $R_v(t_v)$, at the time of crossing it). 
The circularity $\epsilon \equiv J/J_c(E)$ (e.g. Lacey \& Cole 1993) 
is the ratio of the angular momentum $J$ of the orbit to that, $J_c(E)$, 
of a circular orbit with the same energy $E$. Radial orbits have 
$\epsilon = 0$, while $\epsilon = 1$ for circular ones. 
The figure shows that more eccentric orbits come closer to the
cluster's centre, as naturally expected. The mean pericentric distance
is $r_p/R_v(t_v) = 0.38 \pm 0.26$ (1 $\sigma$ of the
distribution). The distribution for 
$\epsilon$ has an average value of $0.53 \pm 0.23$. Almost radial
or almost circular orbits are much less likely than intermediate ones.
The mean amplitude of the 3-dimensional velocity of satellites as
they cross $R_v$ is $v_{sat}(R_v) = 1405 \pm 234$ km s$^{-1}$, 
with 5, 25, 50, 75 and 95 percentiles at 990, 1260, 1510, 1760 
and 2200 km s$^{-1}$ respectively. In terms of the circular 
velocity of the main halo at $R_v(t_v)$, the same velocity
has 5, 25, 50, 75 and 95 percentiles at 0.77, 0.99, 1.12, 1.21 and
1.38 respectively.

We are now in the position to check if satellites only partially 
captured and satellites completely merged have different dynamical 
histories. One possibility is that the formers are on more circular
orbits than the latters, so that only the satellite outskirts are
tidally stripped by the main progenitor, and the rest of the
satellite escapes. At the other extreme, satellites in more 
eccentric or radial orbits would be completely swallowed.
For satellites in these two extreme situations we found average
circularities $\epsilon(f_c<0.2) = 0.55 \pm 0.01$ and 
$\epsilon(f_c>0.8) = 0.527 \pm 0.007$ ($1\sigma$ of the mean, 
now considering satellites at all $t_{id}$). 
Therefore the difference is not significant enough
to explain the effect of Fig.~\ref{fig:fmer}. Another
possible explanation is that satellites with $f_c\ll 1$ 
have already had a first encounter with the cluster, but they 
had enough kinetic energy to emerge from it once, leaving behind 
only part of their mass. The average 3-dimensional
velocities when crossing $R_v$ are, for the same subsets,
$v_{sat}(R_v; f_c<0.2) = 1370 \pm 23$ km s$^{-1}$ and
$v_{sat}(R_v; f_c>0.8) = 1239 \pm 12$ km s$^{-1}$ ($1\sigma$ 
of the mean). 
Here the difference is more significant: indeed satellites 
captured by less than 20 per cent in mass have a higher speed, 
at a $3\sigma$ confidence level, than those completely 
captured. Perhaps a combination of this effect and of the 
former is responsible for the shape of the distribution 
shown in Fig.~\ref{fig:fmer}.

\begin{figure*}
\centering
\epsfxsize=12.truecm\epsffile{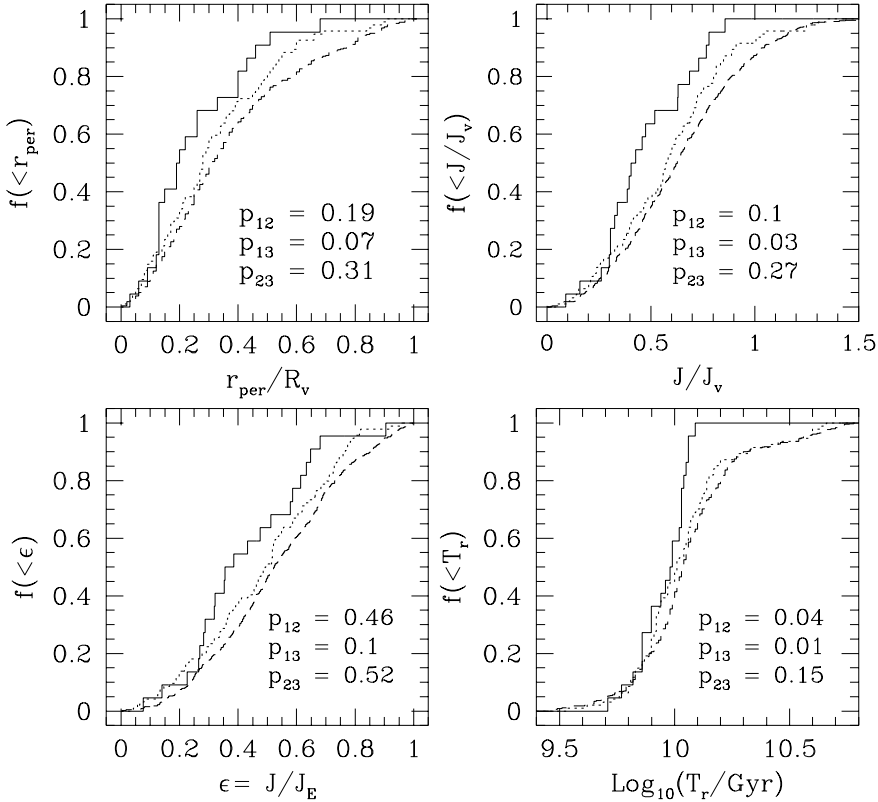}
\caption{Dependence of some orbital parameter on the satellite mass.
Each panel shows the cumulative distibution of a parameter, in 
three mass ranges: $m_v/M_v > 0.05$ (solid lines); 
$0.005 < m_v/M_v < 0.05$ (dotted lines) and 
$m_v/M_v < 0.005$ (dashed lines).
Top Left: orbital pericentre $r_p/R_v$; Top Right: amplitude of the
angular momentum, $J/J_v$; Bottom Left: orbit circularity $\epsilon$;
Bottom Right: radial period $T_r/$Gyr. In each panel, the numbers 
shown are the result of a KS test between every pair of distribution:
$p_{12}$ refers to the solid-dotted pair; $p_{13}$ to the
solid-dashed, and $p_{23}$ to the dotted-dashed.}
\label{fig:orb2}
\end{figure*}

To see if there is any dependence of these and the other orbital
parameters on the mass of the satellite (relative to the main 
cluster progenitor), we binned the data according to the relative 
mass of the satellite, $m_v/M_v$, where $M_v$ is calculated at the 
time $t_v$ when the satellites cross $R_v$.
The results are shown in Fig.~\ref{fig:orb2}, where we plot the
cumulative distributions of $r_p/R_v$, of $\epsilon$, of the amplitude
of the angular momentum of the orbit relative to that of a circular 
orbit at the virial radius, $J/J_v$, and of the radial period $T_r$ 
of the orbit.
Data are binned in mass, in the three intervals $m_v/M_v < 0.005$,
$0.005 < m_v/M_v < 0.05$, and $m_v/M_v > 0.05$.
The figure shows some trend for more massive satellites to be on 
slightly more eccentric orbits, which penetrate deeper in the 
cluster, with lower (specific) angular momentum $J/J_v$ and with 
a shorter
orbital radial period $T_r$. We estimated the significance of these 
trends by a KS test performed on each pair of distributions. The
probability of drawing such distributions by chance, from an
identical underlying population, is given in each panel of the figure.
This test shows that the trend with mass is statistically more 
significant for the radial period $T_r$ and the amplitude of angular 
momentum $J/J_v$, and less for the pericentric distance $r_p/R_v$
and the shape $\epsilon$ of the orbits.
In the next Section we briefly discuss a possible explanation
for these results, although we stress that these trends with mass 
are rather weak.
An even weaker trend was also found for the kinetic energy of 
satellites as they cross $R_v$: more massive satellites approach 
the main halo at slightly lower speed than less massive ones. 
The effect is small, less than 10 per cent for the velocities, and is 
possibly due to dynamical friction.
We finally looked for dependence of these parameters on the mass of
the final cluster, or on the time of identification $t_{id}$,
but did not find any significant trend in the results.

\section{Discussion}
\label{sec:disc}

\subsection{Collapse anisotropy and shape analysis}
\label{sec:anys}
Cosmological $N$-body simulations have shown that, in most scenarios,
the gravitational collapse of matter happens first along pancakes, 
or sheet-like structures. Then filaments form, possibly at the 
intersection of different pancakes. Matter in these filaments 
collapses into dark matter haloes, and these flow along the filaments 
towards the potential minima, where they form galaxy clusters, 
perhaps at filaments' intersections.
The collapse is therefore a highly anisotropic process. 
Recent observations of the distribution of substructure in X-ray 
clusters (West, Jones \& Forman 1995) seem to indicate that 
anisotropic collapse is common also in real clusters. West et al. 
(1995) suggest that this anisotropic formation may be responsible 
for the shape and orientation of clusters with the surrounding 
large-scale structure, on scales $\sim 10 h^{-1}$ Mpc, which is
observed in different samples (e.g. Binngeli 1982; West 1989;
Plionis 1994). A similar alignment is also found in cosmological
$N$-body simulations (West, Villumsen \& Dekel, 1991; Splinter et
al, 1997).
The link between cluster shapes and merging histories has been
addressed in the past (van Haarlem \& van de Weygaert 1993).
It is also recognized that some alignment exists between the
cluster body and its velocity ellipsoid (e.g. Warren et al. 1992).

\begin{figure*}
\epsfxsize=12.truecm\epsffile{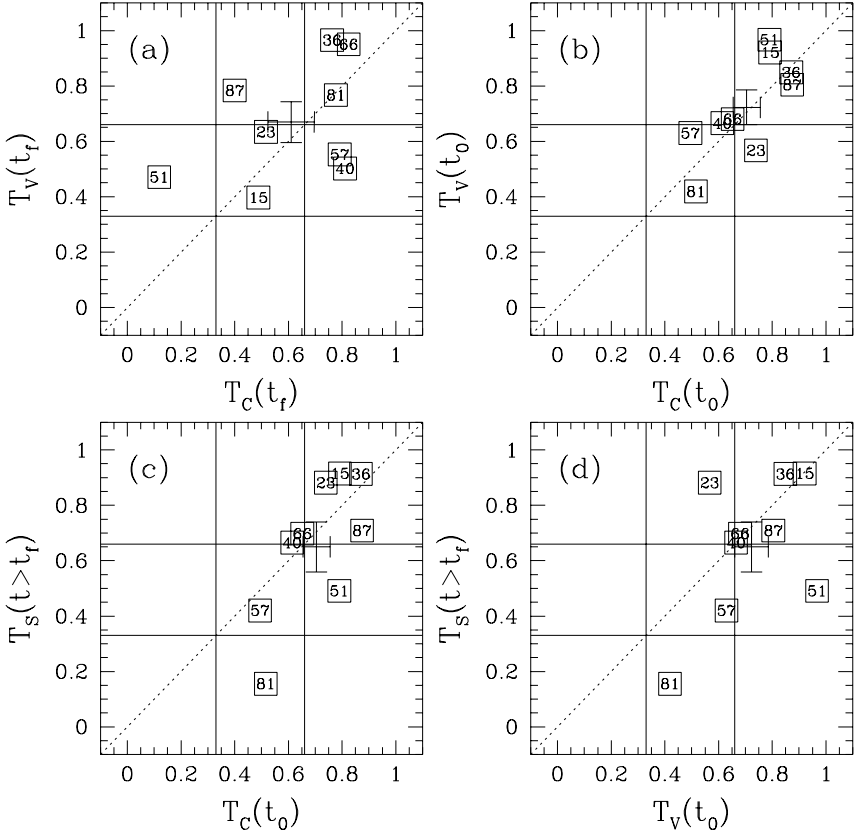}
\caption{Shape Correlations between the cluster (C), its velocity 
ellipsoid (V) and the distribution of its satellites (S). 
Each number labels a cluster.
Panel (a) compares the shape of the cluster and that of its velocity 
ellipsoid at redshift $z_f$ when the main cluster progenitor first 
assembled half of the final cluster mass.
For our simulations $z_f \approx 0.5$.
Panel (b) does the same at $z=0$. Panels (c) and (d) correlates the
spatial and velocity structure of the cluster at $z=0$ with the
distribution of infalling satellites for $z<z_f$.
The grid in each panel separates region corresponding to different 
shapes: oblate ($T<1/3$), triaxial ($1/3 < T < 2/3$) and prolate 
($T>2/3$). The error bars indicate the mean $\pm 1\sigma$ of the mean
for each distribution.}
\label{fig:cosh}
\end{figure*}

The high resolution of our simulations, together with the knowledge 
of their complete merging history, allows us to address the issue
in a more direct and quantitative way than in previous work.
We proceed as follows.
For each cluster of our sample, we first found the principal axes 
of the cluster mass distribution, and of its velocity ellipsoid,
by diagonalizing the mass tensor $I_{ij}$ and the velocity 
dispersion tensor $\sigma^2_{ij}$:
\ba
I_{ij} &=& \sum_k (x_{k,i} - \bar{x}_i)(x_{k,j} - \bar{x}_j), \\
\sigma^2_{ij} &=& \sum_k (v_{k,i} - \bar v_i)(v_{k,j} - \bar v_j),
\ea
where $k$ labels particles and $i,j$ are the Cartesian components 
of the vectors. Since all particles are identical, their mass does
not appear in the expression of $I_{ij}$.
The vector $\bar{\vec x}$ identifies the cluster centre, and the 
mean velocity $\bar{\vec v}$ is calculated using all the particles
inside $0.5R_v$ to limit the bias from infalling substructure.
The sums are extended to all particles within the virial radius. 
Although this choice has been shown to systematically underestimate 
the true axial ratios (Warren et al. 1992), this is not relevant 
for the main purpose of the present analysis.

To characterize the infall pattern we similarly defined the 
mass tensor of the satellite distribution:
\be
\hat{I}_{ij} = \sum_k \hat{x}_{k,i} \hat{x}_{k,j} m_{k,v}.
\ee
Here $\hat{x}$ is the unit vector pointing in the direction 
$(\theta,\phi)$ of approach of the satellite towards the
main cluster progenitor. This direction is measured at time 
$t_{id}$, just before the merging of the two. Each contribution
to $\hat{I}_{ij}$ is now weighed by the satellite's virial mass 
$m_v$. The reference system is centred on the main cluster 
progenitor.

We then compared the shapes of the three ellipsoids and the
relative orientations of the three major principal axes.
Shapes can be classified using the triaxiality parameter 
(Franx, Illingworth \& de Zeeuw 1991):
\be
T = {(a^2 - b^2) \over (a^2 - c^2)},
\ee
where $a \geq b \geq c$ are the three principal axes of the ellipsoid.
We shall call {\em oblate} ellipsoids with $0 < T < 1/3$,
{\em triaxial} those with $1/3< T < 2/3$, and {\em prolate} those 
with $2/3< T < 1$ (Warren et al. 1992), and will label $T_C$, 
$T_V$ and $T_S$ the triaxiality parameters referring respectively 
to the cluster shape, to its velocity dispersion ellipsoid and to 
its satellite distribution as defined above.
We quantify the alignment in the orientation of two ellipsoids 
$E_1$ and $E_2$ using the cosine of the (3 dimensional) misalignment 
angle between the two major axes $a_1$ and $a_2$, defined by 
$\cos(\Delta) = \cos\theta_1\cos\theta_2 + \sin\theta_1\sin\theta_2
\cos(\phi_2 - \phi_1)$. This is uniformly distributed in $[0,1]$
for a random distribution of angles. Values larger than 0.5 
(corresponding to misalignment angles smaller than $60\degr$) 
indicate a correlation in orientation.

In order to study the time evolution of shapes and orientations,
we calculated them at two different times: at redshift $z=0$ and 
at redshift $z_f$ when the main cluster progenitor first assembled 
half of the final cluster mass. 
We only consider satellites identified at $z \leq z_f$. We do so
because for $z > z_f$ the most massive progenitor may be on 
different branches of the merging tree at different redshifts.
In such a case one cannot relate the satellite infall to the
halo shape. Moreover, at early times satellites are identified
at larger distances from the main halo, in units of its virial 
radius, and the direction of infall is less well determined
due to possible perturbations on the orbits.

\begin{figure}
\epsfxsize=\hsize\epsffile{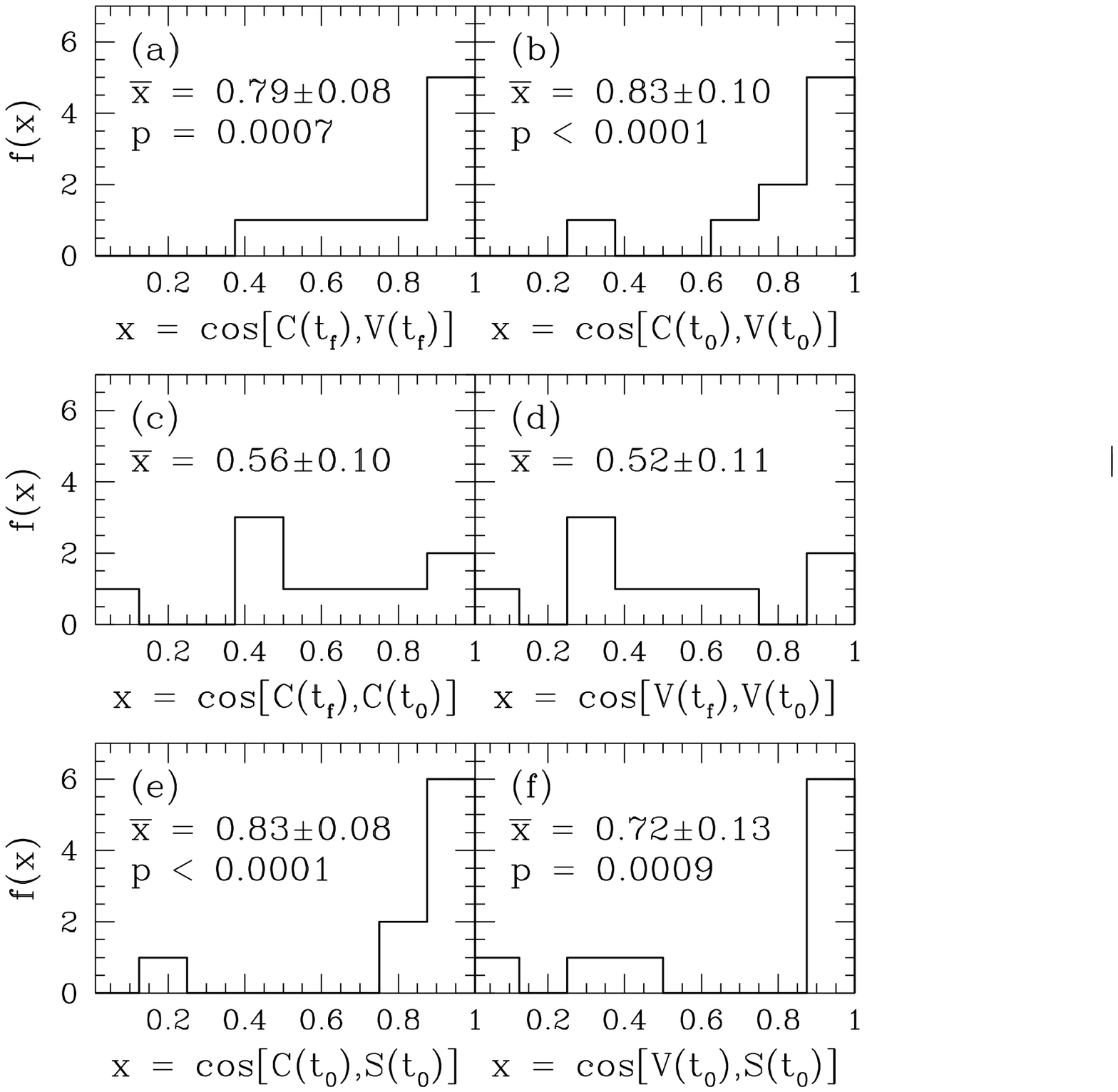}
\caption{Orientation correlations between the major axes of the
cluster shape (C), its velocity ellipsoid (V) and the distribution 
of its satellites (S). Each panel shows the distribution of the
cosine of the misalignment angle for a pair of axes, and is labelled 
with the mean and $1\sigma$ of the mean.
In the relevant cases we also give the probability $p$ of obtaining 
the measured mean alignment from randomly oriented vectors.}
\label{fig:cori}
\end{figure}

The results of our analysis are presented in Figs~\ref{fig:cosh}
and \ref{fig:cori}, and in Table~\ref{tab:1}. They refer to
satellites with $f_c \geq 0.2$.
Fig.~\ref{fig:cosh} correlates, for each cluster of our sample, 
the triaxiality parameter of the cluster shape, that of its velocity 
dispersion and those of the satellite distribution. 
Fig.~\ref{fig:cori} does the same for the cosine of the misalignment 
angle $\cos(\Delta)$.
Table~\ref{tab:1} gives the ratio of major to minor axis for
each considered ellipsoid.

\begin{table}
\caption{Ratio $a/c$ of longest to shortest axis for the cluster
shape ellipsoid (C), for its velocity ellipsoid (V) and for the 
ellipsoid defined by the infall pattern of satellites (S). For each 
ratio, the mean $\pm 1\sigma$ of the mean is given.}
\begin{tabular}{crr}

Label & \multicolumn{1}{c}{$a/c$} & \multicolumn{1}{c}{$a/c$} \\
& \multicolumn{1}{c}{$t_f$} & \multicolumn{1}{c}{$t_0$} \\

\hline
 C  & $1.57 \pm 0.08$  &  $1.60 \pm 0.06$ \\
 V  & $1.45 \pm 0.06$  &  $1.32 \pm 0.05$ \\
 S  &  \multicolumn{1}{c}{-} &  $3.16 \pm 0.30$ \\
\hline

\end{tabular}
\label{tab:1}
\end{table}

From the figures and the table we can make the following comments.
The cluster shape and velocity ellipsoids are preferentially prolate 
or triaxial, and somewhat correlated in shape, at early times
(Fig.~\ref{fig:cosh}a). As evolution proceeds, shapes do not significantly
evolve. Although the correlation improves from Fig.~\ref{fig:cosh}a
to Fig.~\ref{fig:cosh}b, this change is not statistically significant
according to a standard KS test.
Table~\ref{tab:1} shows that the elongation of cluster shapes and 
of their velocity ellipsoids do not appreciably change in time.
A strong shape--velocity correlation is shown in the axes 
orientation, both at early and late times (Fig.~\ref{fig:cori}a,b).
By the present time the mean misalignment between the cluster 
shape and its velocity ellipsoid is of the order of $30\degr$,
with a very high statistical significance.

On the other hand, both the cluster and its velocity ellipsoid 
change orientation in time, so that there is no very significant 
correlation between the cluster orientation at $t_f$ and that 
at $t_0$ (Fig.~\ref{fig:cori}c), and between the corresponding
velocity ellipsoids (Fig.~\ref{fig:cori}d).

The final cluster shape is strongly aligned ($\approx 30\degr$ on 
average) with the infall pattern of satellites accreted after $t_f$ 
(Fig.~\ref{fig:cori}e).
The shapes also are similar (Fig.~\ref{fig:cosh}c).
A similar correlation is observed for the relative alignment of 
the velocity ellipsoid and of the satellite pattern 
(Fig.~\ref{fig:cori}f). The shapes are correlated in that they
tend to avoid oblate configurations (Fig.~\ref{fig:cosh}d). 
The satellite pattern is prolate or triaxial in eight clusters
out of nine. The axial ratio $a/c$ in Table~\ref{tab:1} indicates 
a very anisotropic infall.

The same analysis on the orientation of the minor and intermediate
axes of the ellipsoids gives a picture consistent with these
results. In particular, it appears that the major and intermediate 
axes of an ellipsoid may sometimes switch during cluster evolution, 
causing anticorrelations similar to those observed in 
Fig.~\ref{fig:cori}c,d. 
The minor axis instead remains rather well defined.

To summarize, the present results show that the anisotropic infall 
of matter on the forming cluster is to a large extent responsible 
both for the final cluster shape and for its velocity structure. 
Both are strongly correlated, in shape and orientation, with each 
other and with the infall pattern of merging satellites.
In their analysis, van Haarlem \& van de Weygaert (1993) found that 
the cluster orientation is mainly determined by the last infalling 
subcluster. Although our results are in general agreement with theirs
in highlighting the origin of such orientation, the present analysis
suggests that cluster alignments are stable enough to be observed for 
a long time, as the correlations we found reflect the whole cluster 
history after its formation at $z_f \approx 0.5$. 

Even if we have not directly checked the alignment of our clusters
with the neighbouring large-scale structure, our analysis strongly
suggest this picture, since the infalling satellites are tracers
of the matter distribution around the clusters.
Therefore our results are fully consistent with the alignments 
observed between clusters and the surrounding structure on scales 
$\sim 10 h^{-1}$ Mpc, and favour an explanation of these in terms of 
structure accretion along preferred directions, e.g. along filaments.
The shape and orientation of clusters would then be mainly determined 
by the large-scale initial density field.

This conclusion may also explain the trend in the orbital parameters
shown in Fig.~\ref{fig:orb2}. Massive satellites are more 
likely to reside in dense filaments than less massive ones. 
Therefore, the formers should approach the main cluster 
progenitor along more eccentric orbits than the latters, with
lower specific angular momentum and a closer pericentric distance,
as indeed found.

\subsection{Universal density profiles}
\label{sec:denpr}

\begin{figure*}
\epsfxsize=\hsize\epsffile{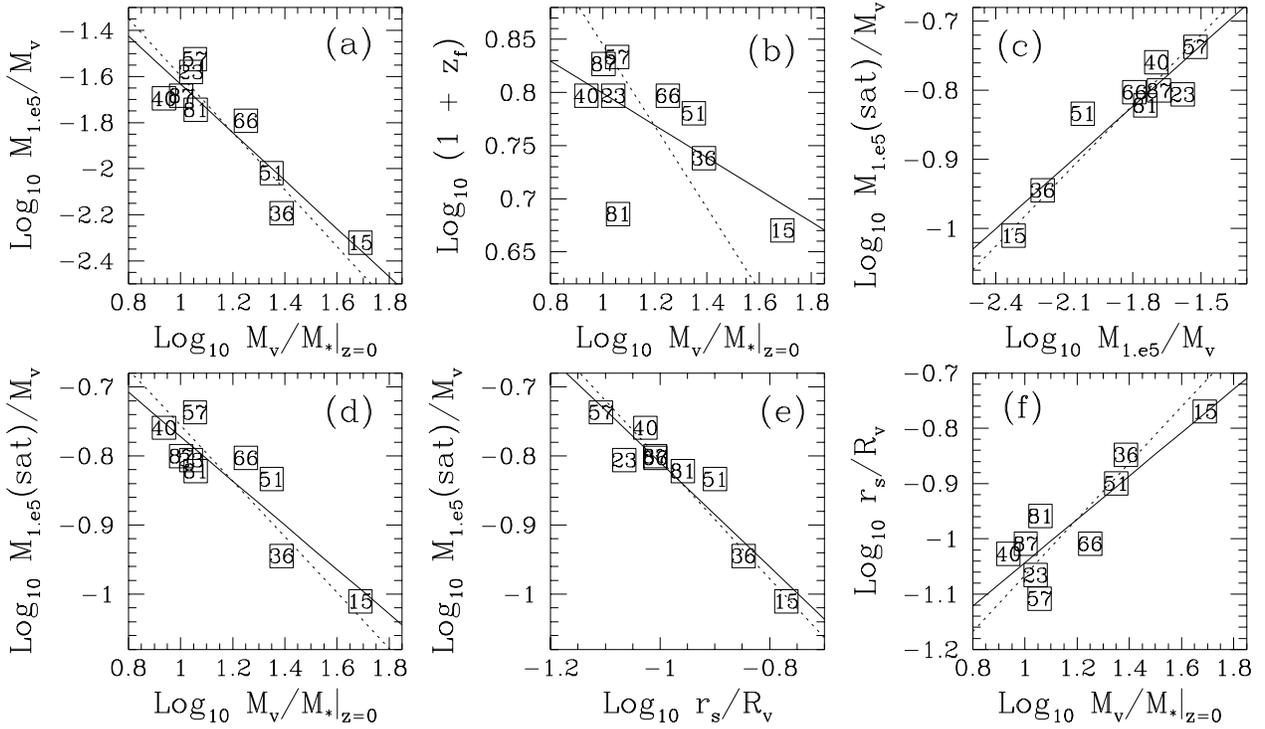}
\caption{Correlations between the cluster virial masses $M_v$, the 
fraction of cluster mass in cluster cores $M_{1.e5}/M_v$, the fraction 
of cluster mass which was in satellite cores $M_{1.e5}(sat)/M_v$,
the cluster formation redshift $z_f$ and the scale parameter $r_s$ 
of the cluster density profile. $M_* = (4/3)\pi R_*^3\rho_b$ is the 
characteristic non-linear mass of the model, with $\rho_b$ the mean
background density of the universe and $R_*$ the scale corresponding 
to a linear matter overdensity $\delta(R_*) = 1.69$.
Each number labels a cluster.
The two straight lines are equal weighted-least square fits of the 
direct (solid) and inverse (dotted) relation. The difference in their 
slope gives an  idea of the goodness of the correlation. 
Definitions of the plotted quantities and details on the figure are 
given in the text.}
\label{fig:ds1}
\end{figure*}

Navarro et al. (1996) have recently found that dark haloes from their 
cold dark matter simulations exhibit remarkable similarity in shape, 
over a wide range of masses. 
After rescaling lengths in units of the halo virial radius $R_v$,
they could fit the dark matter density profile of objects from 
the size of dwarf spheroidals to that of galaxy clusters using 
the one parameter fit 
\be
{\rho(r) \over \rho_b} = {\delta_s \over (r/r_s)(1 + r/r_s)^2}.
\label{eq:nfw}
\ee
The free parameter is the {\em scale radius} $r_s$, which in turn
determines the characteristic halo overdensity $\delta_s$, defined 
by Equation~(\ref{eq:nfw}).
Navarro et al. (1996) found that, expressing halo masses in terms 
of the characteristic non-linear mass $M_*$ corresponding to a 
linear overdensity of order unity, more massive haloes have larger 
scale radii, i.e. larger values of $r_s/R_v$, or equivalently 
have lower characteristic overdensity $\delta_s$. Further work 
(Cole \& Lacey 1996; Tormen et al. 1997, Navarro, Frenk \& White 1996b) 
have confirmed this trend and extended the result to different 
cosmological models.
This behaviour is in fact consistent with the expectations 
of hierarchical clustering, where structures form in a bottom-up 
fashion. Since at earlier times the universe was denser, one 
expects that less massive haloes, which formed earlier, have higher 
characteristic density $\delta_s$, while more massive haloes, 
which form later from the merging of less massive ones, have 
lower values of $\delta_s$.
Navarro et al. (1997) link the characteristic density of a halo 
to the epoch when its progenitors first collapsed. They find a
good agreement between their numerical results and the analytical 
predictions of the extended \ps formalism.
Note however that Navarro et al. define the formation redshift 
of a halo in a way different from ours.

A physical mechanism for the shape of a universal dark matter profile
has been recently proposed by Syer \& White (1996). In their model,
the density profile of a cluster is determined by the properties of 
its merging satellites. In particular, satellites dense enough to 
survive the tidal stripping of the parent cluster remain self-bound 
while they sink to the cluster centre, dragged by dynamical friction, 
and are therefore responsible for the inner shape of the halo density 
profile.
This idea, together with the hierarchical clustering model,
potentially provides a deeper explanation for the empirical 
relation between mass and characteristic density of haloes.
Since we have reconstructed the merging history of all our 
simulated clusters, it is easy to test the idea directly.

\begin{figure*}
\epsfxsize=\hsize\epsffile{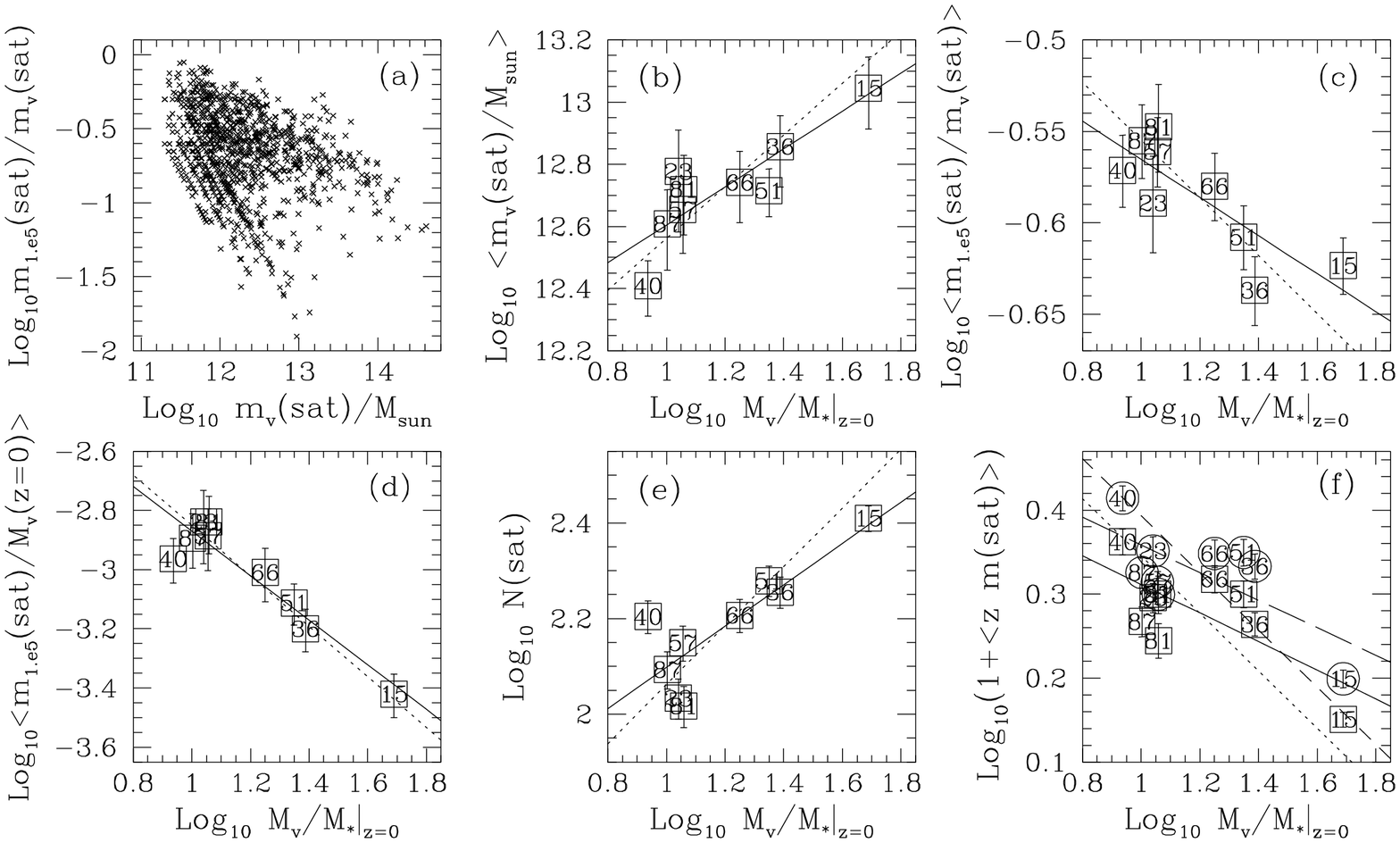}
\caption{Correlations between satellite properties and mass of
the final cluster. The straight lines are equal weighted-least square 
fits of the direct (solid) and inverse (dashed) relation. 
The difference in their slope 
gives an idea of the goodness of the correlation. Each number
label refers to a cluster. Error bars are $\pm 1\sigma$ of
the mean. In panel {\em (f)}, squares are mass weighted by 
$m_v$, the virial mass of satellites. Circles are mass weighted 
by $m_{\delta_t}$, the satellite mass above an overdensity 
threshold $\delta_t = 10^5$. In this panel, the solid and dotted
lines refer to the squares, while long dashed and short dashed
lines fit the circles.}
\label{fig:ds2}
\end{figure*}

We first need a definition to identify the matter in the dense
cores of satellites. For this we take an overdensity threshold 
$\delta_t$, which will always be referred to the mean background 
density at $z=0$. This is important, since we need to compare 
the density of satellites identified at different redshifts
with the density of the final cluster, and we must put all
haloes on equal footing.
For the present analysis we took $\delta_t = 10^5$. 
Each satellite, identified at any redshift, has some fraction 
of its mass above $\delta_t$. This we shall call the mass in
{\em satellite cores}. We shall say that a satellite is {\em denser}
than another if a larger fraction of its total mass is stored
in its core, as just defined. By analogy, the mass above the
threshold  $\delta_t$ in the final clusters: $M_{\delta_t}$, 
will be referred to as mass in {\em cluster cores}.

We show in Fig.~\ref{fig:ds1} some correlations between 
different quantities derived both from the final clusters 
and from their population of satellites. We used satellites
identified at any time, and with $f_c>0.2$.
Looking at the figure, we can make the following statements.
\begin{enumerate}
\item
Less massive clusters are more centrally concentrated
(Fig.~\ref{fig:ds1}a):
the fraction $M_{\delta_t}/M_v$ of cluster mass above the
fixed overdensity $\delta_t = 10^5$ at $z=0$, that is the
fraction of mass confined in the {\em cluster core}, is 
inversely proportional to the total cluster mass $M_v$.
\item
Less massive clusters usually form at higher redshift
(Fig.~\ref{fig:ds1}b): however, the scatter in the
relation is fairly big. The correlation does not improve
if one tries other definitions of formation time, e.g.
changing the fraction of mass required in the most massive
progenitor from the usual value of 50 per cent.
\item
The fraction $M_{\delta_t}(sat)/M_v$ of cluster mass provided
by satellite cores is proportional to the final fraction 
$M_{\delta_t}/M_v$  of mass in the cluster core (Fig.~\ref{fig:ds1}c). 
Therefore, the same fraction $M_{\delta_t}(sat)/M_v$ is inversely 
proportional to the final cluster mass (Fig.~\ref{fig:ds1}d).
\item
The last correlation implies the correlations between 
$M_{\delta_t}(sat)/M_v$ and the scale radius of the profile, 
$r_s/R_v$ (Fig.~\ref{fig:ds1}e). The original correlation
between scale radius and final cluster mass (as in Tormen
et al. 1997, their Fig.~16) is reproduced in Fig.~\ref{fig:ds1}f.
\end{enumerate}
Notice that, while the correlation involving the scale radius
implies a fit to the profile, the others, and in particular that
of Fig.~\ref{fig:ds1}a, is model independent as is directly
measured from the mass profiles.

Let us now focus on the correlation of Fig.~\ref{fig:ds1}d,
and try to identify which factors determine it. Is it
the mass of satellites? Their density? Their merging redshift?
Their number? Or what combination of these?
To answer this, we need to look in more detail at the 
characteristics of the satellite population. These are shown 
in Fig.~\ref{fig:ds2}.
We first note, from Fig.~\ref{fig:ds2}a, that smaller satellites 
are usually denser, that is a higher fraction of their total mass 
comes from their {\em cores}, as defined above. This trend is also 
seen at any fixed time.
Secondly, Fig.~\ref{fig:ds2}b shows that, on average, 
smaller clusters form from smaller (in physical units) 
satellites. The error bars are $\pm 1\sigma$ of the mean 
values. From these two results we may expect that smaller 
clusters form from denser satellites. This is indeed shown 
in Fig.~\ref{fig:ds2}c.
Now, satellites in smaller cluster are smaller in absolute
terms, but they are actually more massive if rescaled to the 
virial mass of the clusters. Therefore, since they are also 
denser, on average each of them brings to the final cluster 
mass $M_v$ a larger contribution of mass from {\em satellite 
cores}. This is clearly shown in Fig.~\ref{fig:ds2}d.
The next panel, Fig.~\ref{fig:ds2}e, shows that smaller
clusters accrete, on average, a smaller number of satellites.
We recall that this is not an effect of numerical resolution,
as each cluster in our sample is resolved by roughly the same 
number of particles, and we consider in all cases satellites 
with at least 8 particles.
The average contribution from each satellite, given in 
Fig.~\ref{fig:ds2}d, times the total number of satellites,
given in Fig.~\ref{fig:ds2}e, give exactly the total contribution
in mass from satellite cores, given in Fig.~\ref{fig:ds1}d.
This shows that the higher density of satellites merging onto 
smaller clusters wins over their slightly lower number.

We finally ask: how relevant to this result is the formation
redshift of the cluster? If an object forms earlier, it
will accrete on average smaller, and denser, satellites.
One way to answer the question is to look at Fig.~\ref{fig:ds1}b:
on average, the largest progenitor of smaller clusters is
assembled earlier. However, the distribution of formation
redshift predicted by the extended \ps formalism is fairly wide, 
and we noted that the 
correlation in Fig.~\ref{fig:ds1}b is not very tight. 
So we see clusters of very different mass which form almost 
at the same time, like clusters labelled 81 and 15, or 40 and 51. 
However, the same pairs of clusters have very different 
values for $M_{\delta_t}/M_v$, and fairly different values 
of $M_{\delta_t}(sat)/M_v$, as well as different scale
radii for their density profiles (Fig.~\ref{fig:ds1}f).
A different way to address the same question is to measure 
the average redshift at which the cluster final mass was 
accreted. This is just the average value of the mass-weighted 
identification redshift of the satellite population: 
$<z_{id} m_v>$, with $m_v$ the satellite virial mass. 
We can also measure the corresponding average redshift weighted
by the mass in satellite cores: $<z_{id} m_{\delta_t}>$.
Both values are shown in Fig.~\ref{fig:ds2}f, versus the final
cluster mass $M_v$. The figure shows roughly the same trend of
Fig.~\ref{fig:ds1}b, as expected, but here the scatter in the
relation is even larger than there. Therefore the redshift of
formation, or the average redshift of the mergers, is less important 
than the intrinsic density of satellites in producing the result 
of Fig.~\ref{fig:ds1}d.

To summarize, we have seen that, compared to more massive clusters,
in less massive ones a higher fraction of the final mass comes
from satellite cores, i.e. is provided by denser satellites.
These satellites are denser because they are smaller, and not 
necessarily because they are accreted earlier. Moreover, their 
cumulative result is due to their density, not to their number.
We repeated the analysis above using different values for the
density treshold $\delta_t$, and found that, in order to reproduce
the result, $\delta_t$ must be larger than the typical density at
the scale radius, but small enough to be accessible by the resolution
of the simulation.

\begin{figure}
\epsfxsize=\hsize\epsffile{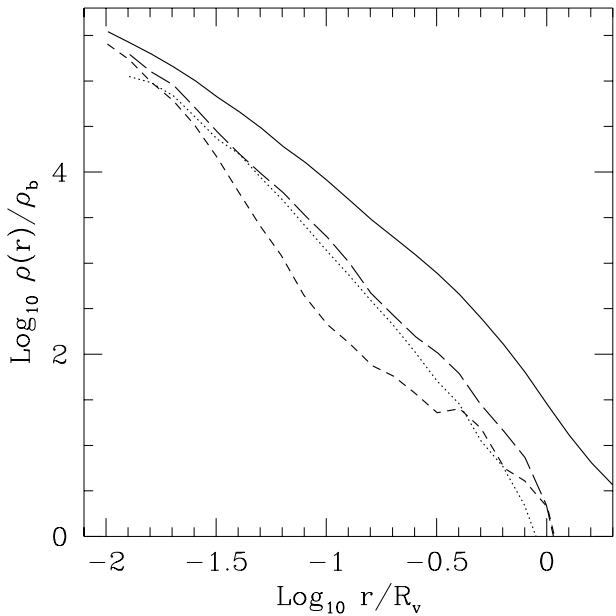}
\caption{Dark matter density profiles at $z=0$. The solid curve
is the profile obtained by using all the particles in the cluster
halo. The other curves were obtained by using only the particles
that were in satellite cores. The dotted line is for satellites
merged at $z>z_f$, the short dashed line for $z<z_f$, the long-dashed
line for all satellites. All profiles are an average over the whole 
cluster sample. Particles from satellite cores are clearly more bound 
than the average particles. Further discussion is given in the text.}
\label{fig:ds3}
\end{figure}

To further show that the mass from satellite cores is directly 
responsible for enhancing the central cluster density, we compare 
in Fig.~\ref{fig:ds3} the density profile of all particles at 
$z=0$ (solid curve) to the density profiles, always at $z=0$, 
calculated using only the particles from satellite cores. 
All profiles are averaged over the cluster sample. 
The dotted curve refers to satellites merged before the formation 
redshift $z_f$, the short-dashed to satellites merged after $z_f$, 
and the long-dashed to all satellites, regardless to their merging 
time.
The figure shows that matter coming from satellite cores is more 
centrally distributed than the average matter in the clusters. 
Among the former, particles merged at earlier times have been heated 
by later mergings, and are less centrally concentrated. Instead, 
particles merged more recently have a higher central density, but 
also a larger spreading at large radii, possibly because they are 
less in equilibrium with the cluster potential.
Quantitatively, the ratio between the half-mass radius of
the mass from satellite cores and that of the whole cluster is
$r_{hm}(sat)/r_{hm}(clu) = 0.45$, $1.12$, $0.72$ for satellites
merged at $z > z_f$, $z \leq z_f$ and at any $z$ respectively.
No difference is found in the velocity structure of the different
particle sets. Therefore, particles coming from the cores of 
satellites are more bound to the final system than other 
particles. This confirms the interpretation of the secondary 
infall picture as a conservation of the binding energy ranking 
(Quinn \& Zurek 1988; Zaroubi et al. 1996).

Finally, the ratio $M_{\delta_t}(sat)/M_v$ is only part of the 
story, and other parameters are likely to be important 
in determining the final structure of the cluster profile. 
For example, clusters labelled 23 and 51, which accrete a similar 
fraction of mass in satellite cores, have very different 
characteristic density and scale radius (Fig.~\ref{fig:ds1}a,d and f). 
However, the present result makes an important link between the 
cluster dynamical history and its final structure, as it shows 
that the sinking of dense satellites in the centre of their parent 
system is a relevant mechanism to giving the cluster its final 
density profile. As far as this mechanism is concerned, less 
massive clusters are more centrally concentrated because a 
larger part of their mass comes from denser satellites.

\section{Summary and Conclusions}

We have analysed the merging of satellites in $N$-body simulations
of a sample of rich galaxy clusters. After identification of all 
cluster progenitors, we selected those haloes infalling directly 
onto the main cluster progenitor. We investigated the mass 
distribution of these {\em satellites}, and derived the main 
parameters for their orbits. Finally, we applied these results 
to two problems relevant to cluster formation.
The first is the relation between the cluster final shape,
its velocity ellipsoid, and the spatial pattern of the infalling
satellites, a link that can explain the observed alignment
of galaxy clusters with the surrounding large-scale structure.
The second is the correlation between mass and characteristic
overdensity of dark matter haloes.

The main conclusions of this paper may be summarized as follows.
\begin{enumerate}
\item
The mass function of satellites, i.e. of dark matter haloes merging 
directly onto the main cluster progenitor, is in very good agreement 
with the predictions based on Monte Carlo merging trees (Lacey \& Cole
1993). Satellites fall onto the main progenitor along orbits with 
average circularity $\epsilon \simeq 0.5$, avoiding radial orbits 
as much as almost circular ones, and have a mean pericentre at 
$r_v \simeq 0.38 R_v$. More massive haloes tend to fall along more 
eccentric orbits, carrying a smaller specific angular momentum and 
reaching farther into the cluster.

\item
The infall of satellites has a very anisotropic distribution.
Its pattern is to a large extent responsible for the final shape 
and orientation of the cluster and of its velocity ellipsoid. 
At redshift $z=0$ the major axis of the cluster mass distribution 
is aligned to the major axis of the cluster velocity ellipsoid to 
$\approx 30\degr$ on average. The cluster is also aligned with 
the ellipsoid defined by the satellite infall to $\approx 30\degr$ 
on average. A similar alignment exists between the satellite infall
and the final velocity ellipsoid of the cluster.

Since this infall reflects the whole cluster history after the 
formation redshift at $z_f \simeq 0.5$, the pattern is stable enough 
to provide an explanation for the observed alignment between galaxy 
clusters and the surrounding large-scale structure, as recently 
suggested by the observations of West et al. (1995).
This result confirms and quantifies the picture, often seen in
$N$-body simulations, of galaxy clusters forming from haloes 
which flow preferentially along filaments. 
Clusters shapes and orientations would then reflect the 
large-scale initial density field.
The same picture provides a possible natural explanation of why 
massive satellite are accreted along more eccentric orbits than 
less massive ones, as we found in our orbital analysis. This is
because they are more likely to reside along dense filaments than 
less massive satellites. Therefore, the filamentary structure 
acts as a sort of focussing rail which drives massive satellites 
towards the cluster.

\item
The properties of the population of satellites infalling onto
the cluster influence also the final shape of its dark matter 
density profile. Lower mass clusters have a larger fraction 
of their total mass supplied by smaller and denser satellites.
This mechanism is to some extent responsible for the correlation 
between the cluster final mass and the characteristic overdensity
of its profile, observed by Navarro et al. (1996).
\end{enumerate}

\section*{ACKNOWLEDGEMENTS}

It is my pleasure to thank Simon White for his insightful comments 
and valuable suggestions to the present work.
Many thanks also to Antonaldo Diaferio, Sabino Matarrese, Julio Navarro, 
Dave Syer, and especially  Cedric Lacey and Lauro Moscardini for helpful 
discussions and comments.
Financial support was provided by an MPA guest fellowship, by an 
EC-HCM fellowship and by the EC-TMR network "Formation and Evolution
of Galaxies". The numerical simulations presented in this paper were
run at the Institute d'Astrophysique de Paris, which is gratefully
acknowledged.

\end{document}